\documentclass{svjour3}                     
\smartqed  
\usepackage{graphicx}
\usepackage{subfig}
\newcommand{\pb}{$\bar{p}$}
\newcommand{\hmol}{H$_2$}
\newcommand{\opp}[1]{\hat{#1}}
\newcommand{\diff}[1]{\textrm{d}#1}
 \journalname{Hyperfine Interactions}
\begin{document}

\title{
Interaction of antiprotons with Rb atoms and a comparison of antiproton
stopping powers of the atoms H, Li, Na, K, and Rb
}


\author{Armin L\"uhr   \and
        Nicolas Fischer   \and
        Alejandro Saenz 
}
%
%
\institute{
              Institut f\"ur Physik, AG Moderne Optik, Humboldt-Universit\"at
              zu Berlin, Hausvogteiplatz 5-7, D-10117 Berlin, Germany. \\
              Tel.: +49-(0)30-2093-4814\\
              \email{Armin.Luehr@physik.hu-berlin.de}            
}

\date{Received: date / Accepted: date}

\maketitle

\begin{abstract}
\label{txt:abstract}
Ionization and excitation cross sections as well as
electron-energy spectra and stopping powers of the alkali metal atoms Li, Na,
K, and Rb colliding with antiprotons were calculated using a time-dependent
channel-coupling approach. An impact-energy range from 0.25 to 4000\,keV was
considered. The target atoms are treated as effective one-electron systems
using a model potential.  
The results are compared with calculated cross sections for
antiproton-hydrogen atom collisions.  
\end{abstract}
%
%
%
\section{Introduction}
\label{sec:intro}
The motivation for studying antiproton (\pb ) collisions is manifold and only
a selection of motives shall be mentioned here. 
First of all, there is a general interest in collisions with exotic particles. 
Second, \pb\ scattering is a fundamental collision system due to the
fact that the projectile is heavy and has a negative charge resulting in several
advantages. For example, antiprotons do not capture electrons leading
essentially to a one-center problem. They approximately follow a classical
path down to velocities slower than the Bohr velocity. And even
very slow \pb\ are still able to ionize making the investigation of
``adiabatic'' collisions possible.   
Third, a comparison of results obtained from \pb\ and $p$ scattering yields
differential information on the collision process. 

The present motivation originates also from other aspects. One is the facility
design for the Facility for Antiproton and Ion Research (FAIR) including the
Facility for Low-energy Antiproton and Ion Research (FLAIR). In order to set the
requirements of the (low-energy) \pb\ storage rings correctly the influence of
residual-gas atoms and molecules on the \pb\ beam has to be known. Therefore,
quantities like the stopping power or the diffraction of the \pb\ beam due to
collisions are of interest. Furthermore, a theoretical data base of \pb\ cross
sections for various atomic and molecular targets should be provided in view of
future experiments at the Antiproton Decelerator (AD) at CERN and also under
improved conditions at the upcoming FLAIR. 

During the last decades advances have been achieved in the understanding of
\pb\ collisions with the simplest one- and two-electron atoms H and
He. However, in the case of \pb\ + He experiment and theory did not agree for
impact velocities below the mean electron velocity of the ground state for
more than a decade stimulating a vivid theoretical activity. The
discrepancy has been partly resolved by very recent measurements on \pb\ + He
by Knudsen {\it et al.} \cite{anti:knud08}. 

On the other hand, the literature on \pb\ collisions with alkali-metal atoms
is sparse. No measurement has been performed so far and only two calculations
for Li and Na targets by Stary et al.~\cite{anti:star90} as well as by McCartney
and Crothers~\cite{anti:mcca93}, and recently results for Li, Na and K
atoms \cite{anti:luhr08} were presented. This is in contrast to the fact, that
alkali-metal atoms are in principal theoretically and experimentally
feasible due to their electronic shell structure. All core electrons fill
inner closed shells while a single valence electron occupies an outer s
($l=0$) state. 

In this work results for ionization and excitation in \pb\ collisions with Rb
atoms are discussed and compared to previous findings for the alkali-metal
atoms Li, Na and K \cite{anti:luhr08} as well as to calculations with an
atomic H target. Electron-energy spectra are determined and are exemplarily
shown for Rb. Furthermore, electronic stopping powers are presented
for \pb\ collisions with Li, Na, K and Rb. 
The used method has already been discussed elsewhere in some detail 
\cite{anti:luhr08}. Therefore, only the major
approximations, the description of the target atoms, as well as the general
concept are briefly reviewed in the following section.
Atomic units are used unless 
otherwise stated. 
%
%
%
\section{Method}
\label{sec:method}
%
%

The collision process is considered in a semi-classical way using the impact
parameter method. Thereby, the electrons of the target atoms are treated
quantum mechanically while the projectile moves on a straight classical
trajectory ${\bf R}(t)={\bf b} + {\bf v} t$ given by the impact parameter
${\bf b}$ and the velocity ${\bf v}$ which are parallel to the $x$ and $z$
axis, respectively and $t$ is the time.

An effective one-electron description $\Psi(r,t)$ of the collision process is
used
     \begin{equation} 
       \label{eq:tdSE} 
          i \, {\frac{\partial}{\partial t}} \, \Psi({\bf r},t) =  
          \left (\,  \opp{H}_0 + { \frac{-Z_p}{\left| \mathbf{r -
                    R}(t) \right|} + \frac{Z_p}{|\mathbf{R}(t)|}}\,\right )
           \Psi({\bf r},t)\ , 
      \end{equation} 
where $\mathbf{r}$ is the electron coordinate, $Z_p$ is the charge of the
projectile, and $\opp{H}_0$ is the time-independent target Hamiltonian. $\Psi$
is expanded in eigenstates $\phi_j$ of $\opp{H}_0= -\frac{1}{2}\, \nabla^2
+ \opp{V}_{\rm mod}(r)$ centered on the target nucleus.  The model potential
$V_{\rm mod}$ used for the effective one-electron description of the
alkali-metal atoms was proposed by Klapisch \cite{aies:klap71} and the
potential parameters are take from \cite{aies:magn99a}.  The radial part of
the $\phi_j$ is expanded in B-spline functions while the angular
part is expressed in spherical harmonics.

The expansion $\Psi=\sum_j c_j(t)\, \phi_j $ leads to
coupled differential equations for the expansion coefficients $c_j(t)$ for
every trajectory, i.e., for every $b$ and impact energy $E=M_p\,{\bf v}^2 /\,2$ with
$M_p$ being the antiproton mass. The cross
section $\sigma_j$ for an explicit $E$ for a transition from the initial state
to $\phi_j$ is given by   
\begin{equation} 
  \label{eq:cross_section} 
 \sigma_{j} = 2\, \pi\, \int |\,c_{j}(t\rightarrow \infty,b)\,|^{\,2}\,\,b\;
                         \diff{b}\,,      
\end{equation} 
where the cylindrical symmetry of the collision system has been used. The
electronic energy loss cross section is then given by  
\begin{equation} 
  \label{eq:stopping_power} 
  S =  \sum_j (\epsilon_j - \epsilon_1)\,\sigma_j \,,   
\end{equation} 
where $\epsilon_1$ and $\epsilon_j$ are the energies of the ground and the
$j$-th state, respectively. The cross sections for ionization and excitation
are given by the sum of all $\sigma_j$ in Eq.~(\ref{eq:cross_section}) for
transitions into states $\phi_j$ with positive and negative energies
$\epsilon_j$, respectively.

%
%
%
\section{Results}
\label{sec:results}
The method was extensively tested by studying its convergence behavior 
and, additionally, by detailed comparisons of proton ($p$) cross sections
obtained with the same approach with literature results
\cite{anti:luhr08}. Also, a detailed comparison to the two already mentioned
works \cite{anti:mcca93,anti:star90} which address \pb\ collisions with Li and
Na can be found in \cite{anti:luhr08}.  Furthermore, the present method was
successfully applied to the calculation of collisions including an \hmol\
target treated as an effective one-electron system
\cite{anti:luhr08a,dia:luhr08}. Therefore, no further testing of the method is
presented here.  
%
%
%
\subsection{Ionization}
\captionsetup[subfloat]{labelfont={bf},labelformat=simple}
\begin{figure}
\centering
  \subfloat[Ionization cross section $\sigma_{\rm ion}$ for \pb\,+\,Rb
  collisions  compared with Li, Na, K, and H \cite{anti:luhr08} targets as 
  function of the impact energy $E$. 
  Solid curve, Li; dash--dotted  curve, Na; 
  dashed curve K; double-dash--dotted  curve, Rb;
  dash--doubly-dotted curve, H.
  \label{fig:ionization}]
  {\includegraphics[width=0.48\textwidth]{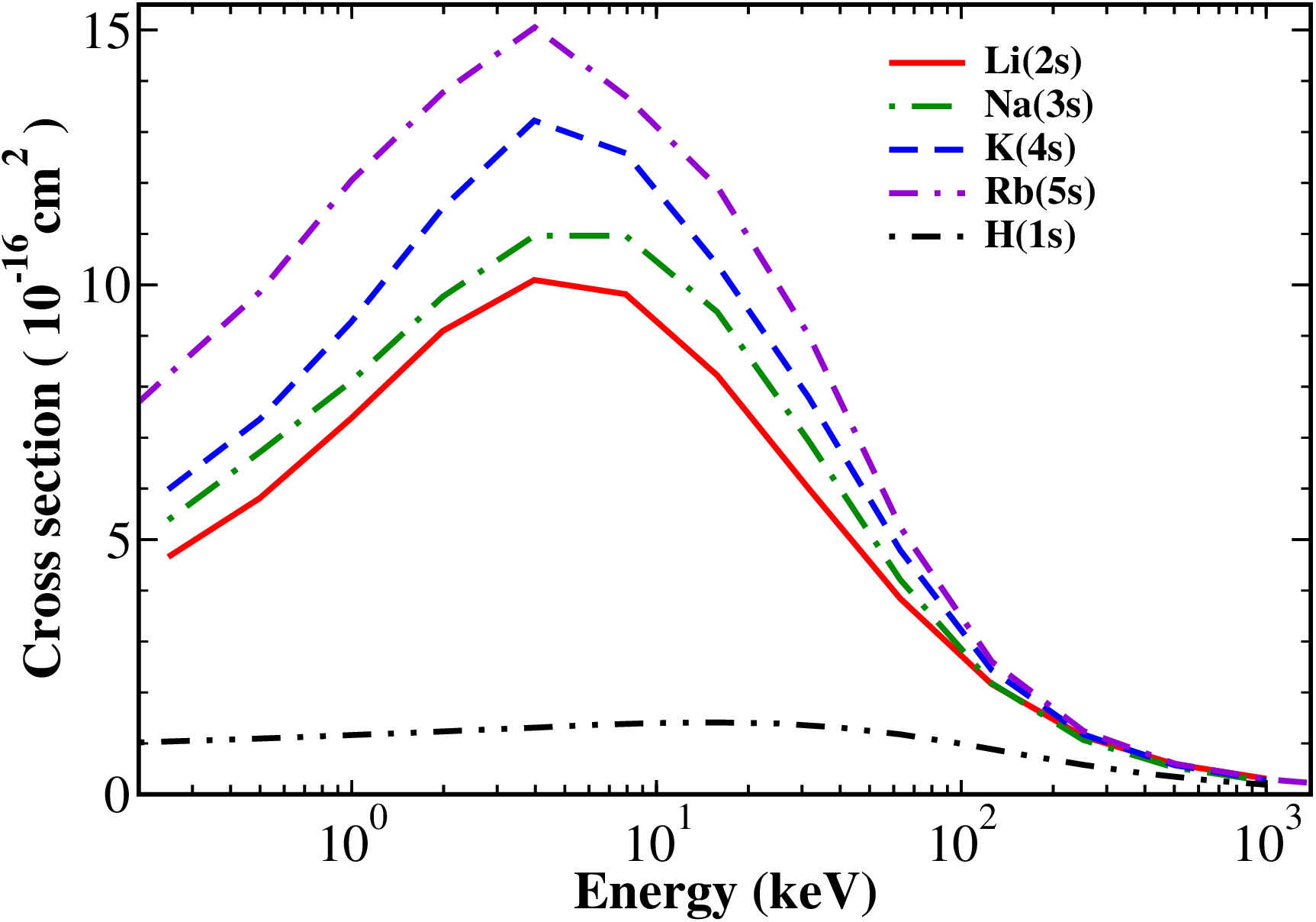}}    
  \hspace{0.02\textwidth}
  \subfloat[Excitation cross section $\sigma_{\rm ex}$ for \pb\,+\,Rb 
  collisions  compared with Li, Na, K, and H \cite{anti:luhr08} targets as 
  function of the impact energy $E$. 
  Solid curve, Li; dash--dotted  curve, Na; 
  dashed curve K; double-dash--dotted  curve, Rb;
  dash--doubly-dotted curve, H.
  \label{fig:excitation}] 
  {\includegraphics[width=0.48\textwidth]{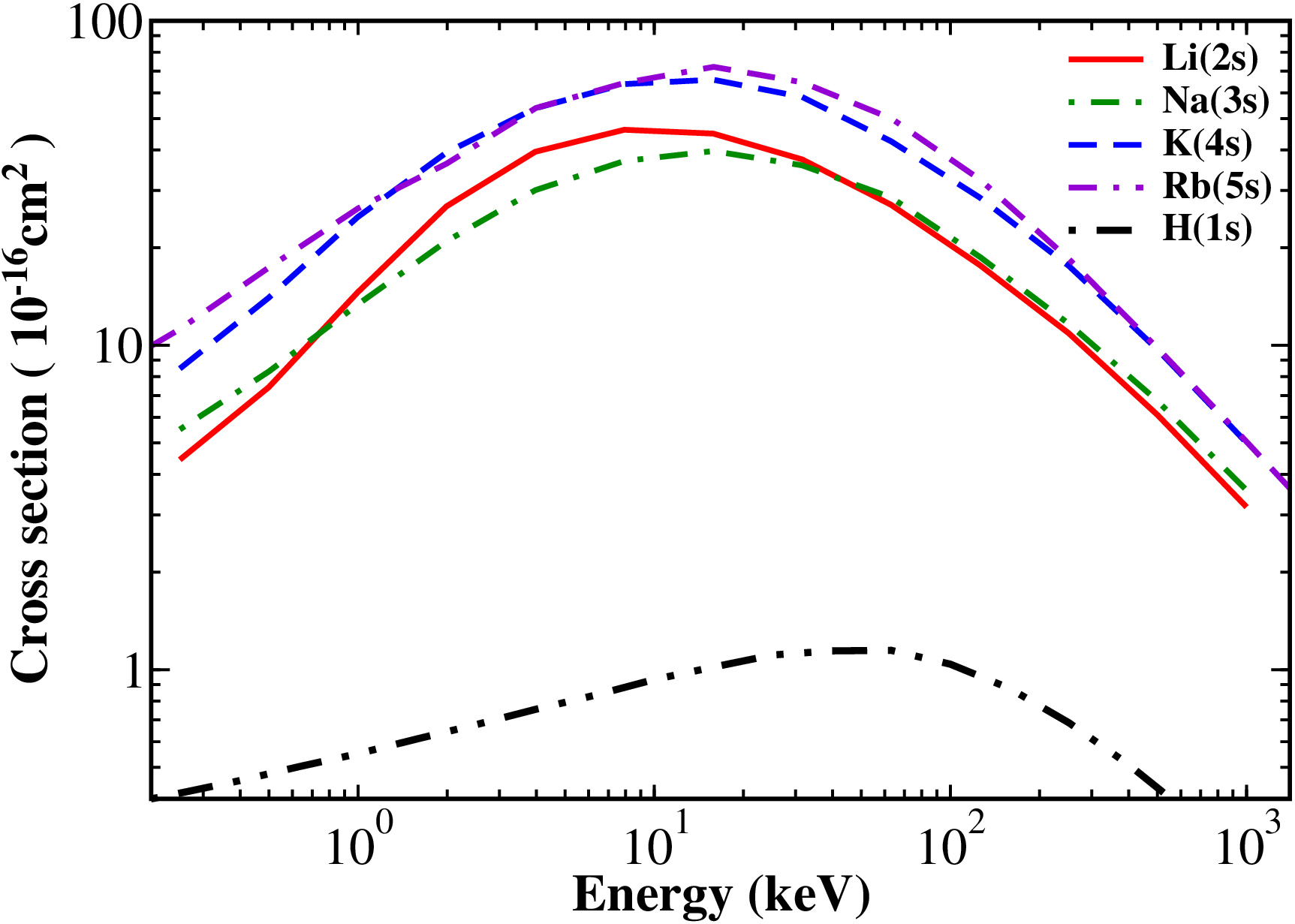}} 
\end{figure}
\addtocounter{subfigure}{2}
The ionization cross sections for \pb\ collisions with Rb are compared in
\subref*{fig:ionization} to the results for Li, Na, K and H
\cite{anti:luhr08}. It can be seen that the curves for all alkali-metal atoms 
share the same qualitative behavior. They differ, however, in
the magnitudes of the maxima which are order according to the chemical element. 
The larger the group number in the periodic table 
the higher the maximum of the ionization cross section. While this ordering is
also prominent at low energies the differences between the atoms vanish at high
impact energies. 

For comparison also the well known ionization cross sections for \pb\ + H
collisions are given in \subref*{fig:ionization}. Similar results for H might
be expected since it is in the same group in the periodic table as the
alkali-metal atoms. However, there are also obvious differences. Namely, the
magnitudes of the cross sections for the alkalis and the H atom differ
considerably and only become comparable at high impact energies $E \ge
500$ keV. Also the position of the maximum for H is shifted from $\approx$\,
5 keV for the alkalis to $\approx$\,25 keV for H.  
The differences of the magnitudes of the ionization curves among the alkalis
as well as between the alkalis and the H atom  in
\subref*{fig:ionization} can be explained with the different ionization
potentials of the target atoms \cite{anti:luhr08}.

A comparison to the results for $p$ collisions is not shown here. However,
it should be mentioned that the findings for $p$ differ considerably from
those for \pb\ for energies $E<100$ keV \cite{anti:luhr08}. For these
energies the electron loss in the case of $p$ impact is much larger than for
\pb\ collisions. This is due to the fact that electron capture becomes the
dominant loss process at low $p$ impact energies which is excluded for \pb .

%
%
%
\subsection{Excitation}

In \subref*{fig:excitation} the excitation cross sections for \pb\ 
collisions with Rb are compared to the results for Li, Na, K and H
\cite{anti:luhr08}. Again, all curves for the alkali-metal atoms
share the same qualitative behavior but differ in the magnitudes of their
maxima. The maxima are, however, not ordered by chemical element. Here, Li
lies above Na. The maxima of the curves are at higher impact energies --- Na, K,
Rb\,$\approx$\,15 keV and Li\,$\approx$\,8 keV --- compared to ionization. In
absolute values all curves lie clearly above the results for
ionization in \subref*{fig:ionization} whereas the curves for Li and Na as
well as for K and Rb are close to each other.
On the other hand, the slope of the curve for Li is more similar to that of K
and that of Na more similar to that of Rb. 

A comparison to the results for excitation of the H atom where the maximum
is around 50 keV shows that the difference is even larger than in the case
of ionization. The differences for excitation among the alkalis
as well as between the alkalis and the H atom can be explained with the
different excitation energies, especially for the 
lowest excited p state ($l=1$) which is the most prominent transition at high
and intermediate $E$ \cite{anti:luhr08}. The excitation energy for the 1s
-- 2p transition in H with 0.375 a.u.\ is in particular large compared to, e.g,
the 5s -- 5p with 0.058 a.u.\ for Rb.

A comparison of the \pb\ excitation results to $p$ collisions with alkalis
yields, in contrast to ionization, no large differences \cite{anti:luhr08}. This
could have been expected since no additional excitation mechanism for one of
the two projectiles exists.  However, smaller differences for \pb\ and $p$
impact exist and will be discussed elsewhere.   

%
%
\subsection{Electron-energy spectra}
%
%
%

\begin{figure}
\centering
%
  \subfloat[%
  Electron-energy spectra for \pb\ + Rb collisions as a function of the
  energy $\epsilon$ of the emitted electron. Spectra are shown for seven
  different \pb\ impact energies (keV): 1, 4, 16, 64, 250, 1000, and 4000. The
  inset shows the region of small $\epsilon$  enlarged.
  \label{fig:spectrum}   ] 
  {\includegraphics[width=0.48\textwidth]{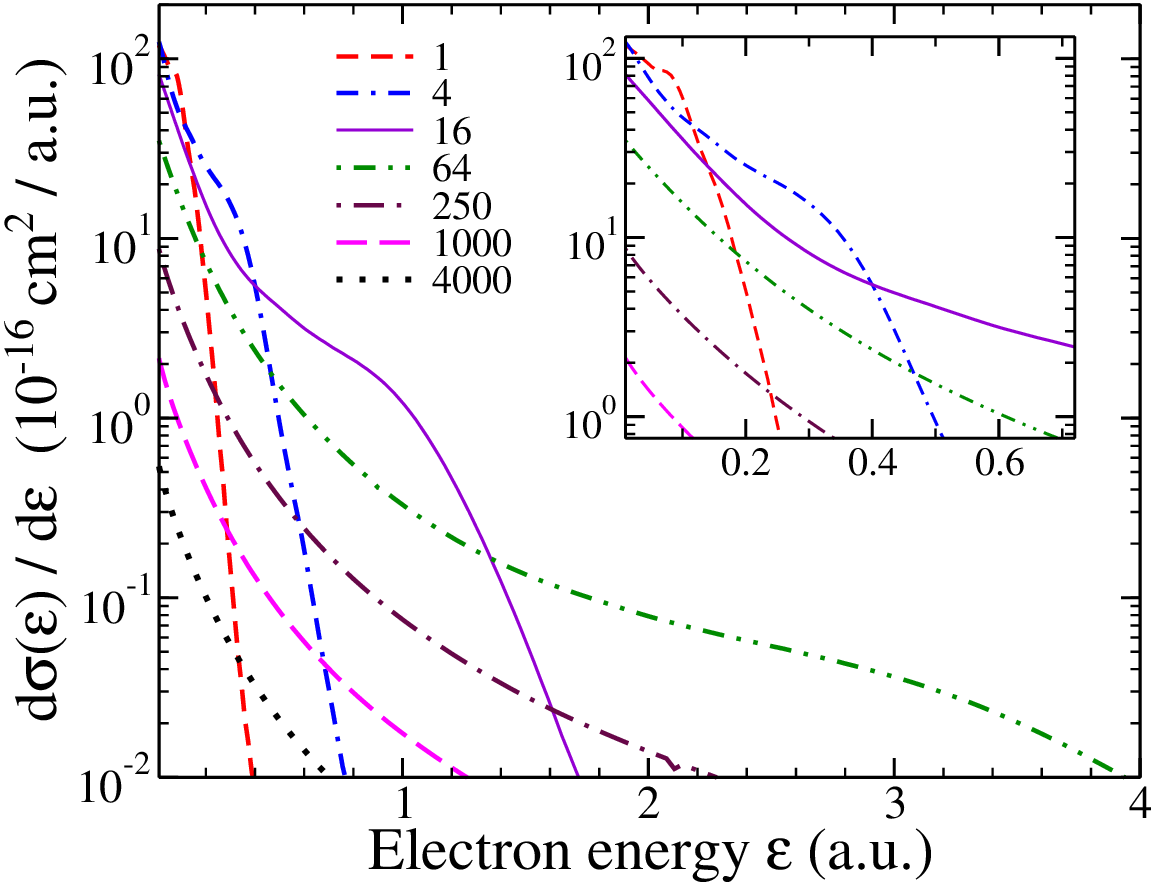}} 
  \hspace{0.02\textwidth}
  \subfloat[%
  Energy loss cross section for \pb\ collisions with Li, Na, K, Rb and
  H as function of $E$. 
  Solid curve, Li; dash--dotted curve, Na; 
  dashed curve K;  double-dash--dotted curve, Rb; 
  dash--double-dotted curve, H; 
  triangles, H, Grande and  Schiwietz~\cite{anti:gran97}. 
  \label{fig:loss} ]  
  {  \includegraphics[width=0.465\textwidth]{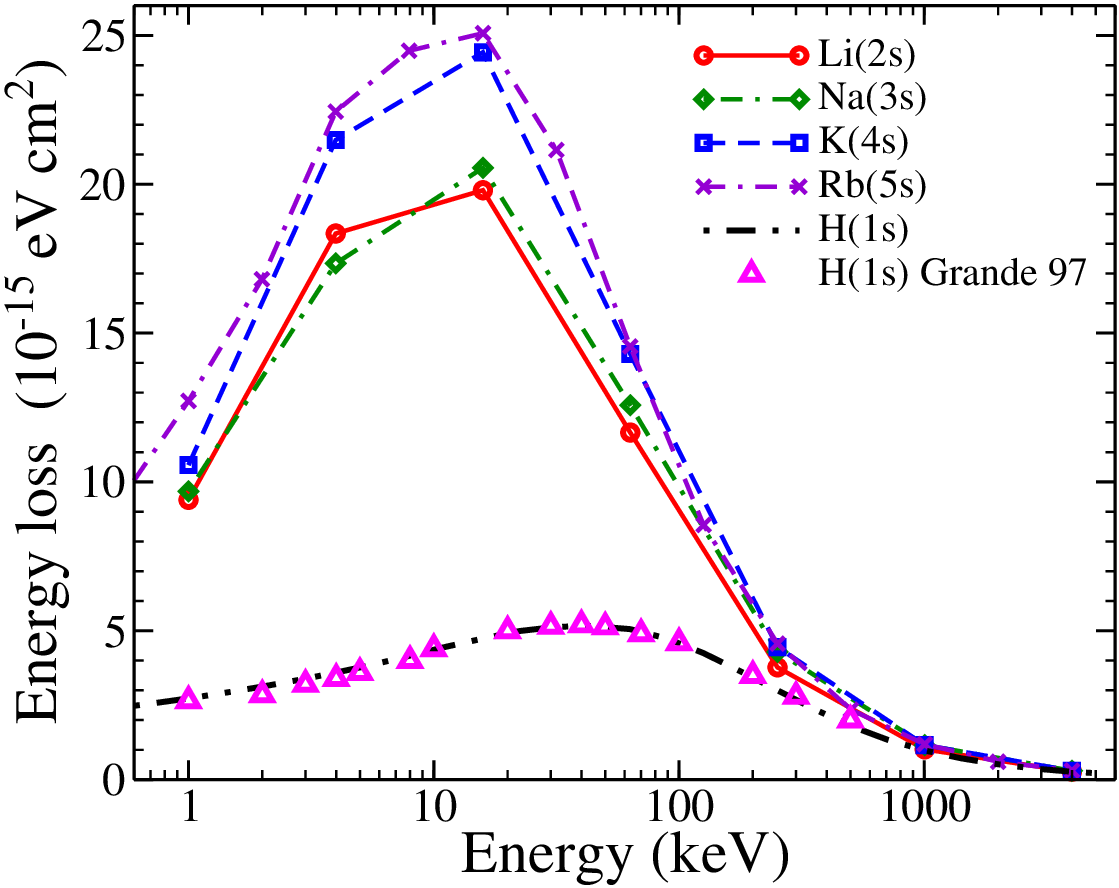}
  }  
\end{figure}
\addtocounter{subfigure}{2}
Besides total cross sections for ionization and excitation also differential
information can be extracted from the collision process. In 
\subref*{fig:spectrum} the electron-energy spectrum, i.e., the cross section
that an electron is emitted with the energy $\epsilon$ is presented for a Rb
atom target. The spectra of the other alkali-metal atoms Li, Na, 
and K  were also determined. They are qualitatively comparable to that of Rb
and are therefore not shown here. Spectra for seven different \pb\ impact
energies $1\le E \le 4000$ keV are given. All curves are smooth, fall off
with increasing $\epsilon$ and no resonance structures can be seen for \pb\
impact which is also the case for \pb\ + \hmol\ collisions
\cite{anti:luhr08a}. This is in contrast to the spectra for electron loss in
the case of $p$ collisions where a pronounced resonance can be observed for
$\epsilon = E / M_p = {\bf v}^2/2$ \cite{anti:luhr09b}. The resonance originates
from the electron capture process by the $p$. Thereby, the ionized and then
captured electron moves with approximately the velocity ${\bf v}$ of the $p$.  

For lower impact energies $E$ the maxima of the spectra at $\epsilon
\rightarrow 0$ are higher and the decrease of the curve with increasing
$\epsilon$ is steeper. For small electron energies   $\epsilon
\rightarrow 0$ the curves are ordered according to their impact energy
$E$. For larger electron energies $\epsilon > 0$ the spectra all share the
same behavior. Always the uppermost curve crosses all lower-lying curves
belonging to larger $E$. This is nicely illustrated in the inset in
\subref*{fig:spectrum}; first for the spectrum for $E$\,=\,1 keV and then for
that for $E$\,=\,4 keV. Consequently, the mean kinetic energy of the ionized
electrons $\bar{\epsilon}$ increases for larger impact energies $E$ of the \pb
. 
%
%
%
\subsection{Stopping power}
The obtained differential information on ionization and excitation can be used
for the determination of the cross section for electronic energy loss also
referred to as stopping power $S(E)$. 
Results for \pb\ collisions with Li, Na, K, Rb, and H are shown in
\subref*{fig:loss}. A comparison of the present findings for \pb\ + H with an
atomic orbital calculation by Grande and Schiwietz \cite{anti:gran97} yields
good agreement and confirms therefore the present implementation.

As could have been expected from the results for ionization and excitation the
electronic energy loss for H targets is smaller than for the alkali-metal
atoms. On the other hand, the difference of $S(E)$   between H
and alkali-metal atoms is, especially for high $E$, not very large. This can
be understood 
regarding Eq.~(\ref{eq:stopping_power}) where the $\sigma_j$ are weighted
with the excitation energies ($\epsilon_j - \epsilon_1$) which are clearly
larger for excitations of an H atom. The differences for Li and Na observed for
ionization and excitation compensate each other for the electronic energy
loss. Also, the results for K and Rb are very similar. In general, it is
possible to conclude that in the case of \pb\ impact the electronic stopping
power of alkali-metal atoms is dominated by the excitation process. Therefore,
a large stopping power can be found for all alkali-metal atoms at $E\approx
15$ keV in a narrow impact energy range when considering the energy loss on a
linear $E$ scale.   

%
%
%
\section{Outlook}
\label{sec:outlook}
Besides the electron-energy spectrum also the angular distribution of the
ionized electrons can be analyzed in order to obtain doubly-differential cross
sections since the wave function $\Psi$ is fully known (at every time
step). Currently, the method is extended to treat also molecular 
targets. Calculations for \hmol\ using a one-electron model potential of the
target have already been performed \cite{anti:luhr08a,dia:luhr08}. A further
development including a full two-electron description of the collision 
process is in progress which considers also molecular properties. Once a
two-electron description is implemented it can be used for atomic targets,
too.  In the limit of vanishing internuclear distance, e.g., He atoms could be
used in order to test the implementation. But also calculations for, e.g.,
alkaline earth metals can easily be done, if again model potentials are used.

\begin{acknowledgements}
The authors are grateful to BMBF (FLAIR Horizon) and {\it Stifterverband f\"ur
  die deutsche Wissenschaft} for financial support.
\end{acknowledgements}

\bibliographystyle{spmpsci}      


%
%

\end{document}